\documentclass[%
 aip,
 amsmath,amssymb,
 reprint,%
]{revtex4-2}

\usepackage{graphicx}
\usepackage{dcolumn}
\usepackage{bm}

\usepackage[utf8]{inputenc}
\usepackage[T1]{fontenc}
\usepackage{mathptmx}

\begin{document}

\preprint{AIP/123-QED}

\title[Measurement of ion displacement via RF power variation for excess micromotion compensation]{Measurement of ion displacement via RF power variation for excess micromotion compensation}

\author{Ryoichi Saito$^{1,2}$}
\email{r\_saito@ee.es.osaka-u.ac.jp}
\author{Kota Saito$^1$}%
\author{Takashi Mukaiyama$^{1,2}$}%
\email{muka@ee.es.osaka-u.ac.jp}
\affiliation{
$^1$Graduate School of Engineering Science, Osaka University, 1-3 Machikaneyama, Toyonaka, Osaka 560-8531, Japan\\
$^2$Quantum Information and Quantum Biology Division, Institute for Open and Transdisciplinary Research Institute, Osaka University, Osaka 560-8351, Japan.\\
}%

\date{\today}

\begin{abstract}
We demonstrate a method of micromotion minimization of a trapped ion in a linear Paul trap based on the precision measurement of the ion trapping position displacement due to a stray electric field in the radial plane by ion fluorescence imaging.
The amount of displacement in the radial plane is proportional to the strength of a stray electric field. Therefore,
we evaluated the micromotion compensation condition by measuring the ion displacements from the ion equilibrium position using two different radial trap frequencies
with various combinations of the compensation voltage.
The residual electric field uncertainty of this technique reached a few volts per meter.
This compensation technique does not depend on the orientation of the incident cooling laser or the detuning and imaging direction. Therefore, this method is suitable for a planar ion trap, a stylus ion trap, which limits the propagation angle of lasers, or miniaturized  ion trap systems for sensing and metrological applications.
\end{abstract}

\maketitle

%

\section{\label{sec:level1}Introduction}
A laser-cooled ion stored in an ion trap is a valuable platform for developing devices based on quantum phenomena.
A single to several trapped ions in a vacuum, considered as a well-isolated system, is used in the development of quantum techniques,
especially quantum computing, quantum simulation, and quantum sensing.
The design and architecture of an ion trap have been investigated to optimize scaling up the number of entangling ions or enable miniaturization.
Examples of ingenious ion traps for quantum devices include the segmented ion trap\cite{Kielpinski2002}, microfabricated ion trap\cite{Stick2006, PhysRevLett.96.253003}, planar trap\cite{Chiaverini, PhysRevA.73.032307}, and stylus trap\cite{Maiwald2009, doi:10.1063/1.4817304}.

An ion in a Paul trap has a motional mode directly driven by the RF electric field that confines the ion in a harmonic potential in terms of pseudopotential approximation,
so-called micromotion\cite{doi:10.1063/1.367318}, regardless of the complexity of the trap architecture.
Because micromotion minimization has a strong relationship with the ion trap's performance,
micromotion compensation is a necessary and essential initial procedure for utilizing trapped ions for precise quantum level manipulation.
Therefore, several compensation methods have been developed. 
These compensation procedures are characterized by an indicator of excess micromotion, and include 
motional sideband spectroscopy\cite{Schulz_2008, doret2012controlling}, RF-photon correlation\cite{doi:10.1063/1.367318, pyka2014high},
parametric excitation\cite{doi:10.1063/1.3665647, tanaka2012micromotion}, atomic loss due to atom-ion collisions\cite{doi:10.1063/1.4809578}, and
ion trajectory\cite{PhysRevA.92.043421}.

This paper demonstrates a micromotion minimization method based on ion position detection with two different trap confinement variations.
In an ideal case, an ion in a Paul trap is stored at the RF null point due to the trap RF electric field, while in general, an unexpected stray electric field shifts the equilibrium position from the RF null point.
The strength of the stray electric field can be evaluated from the measurement of the ion displacement due to two different radial confinements.
Minimization of the ion displacement gives us the micromotion well-compensated codition.
Finally, we obtain a residual electric field uncertainty of a few volts per meter by our compensation measurement.

The method introduced in this study does not depend on the orientation of the incident cooling laser or its detuning and imaging direction.
Therefore this technique is suitable for various ion trap architectures, including planar traps, and miniaturized ion trap systems for metrological applications,
which have structural limitations for an incident laser.
Furthermore, this method does not require any stable lasers to excite the narrow transition for sideband spectroscopy or
any high-speed measuring instruments to detect photon emission timing for RF-photon correlation measurements.
Only a conventional ion fluorescence imaging system is required to determine the ion trapping position.
By comparison with the compensation method of ion trajectory \cite{PhysRevA.92.043421}, which requires continuous scanning of the RF amplitude,
our method is simple and convenient
because we seek the RF null from two different RF amplitudes.
While this compensation method relies on ion fluorescence images,
the residual electric field's limitation is carried out by the resolution of ion images in principle.
We also mention systematic error due to forces other than the electric field;
thus, our imaging resolution of ion position sensing reached a sensitivity of 125 $\rm zN$.
This is comparable to the scattering force of the cooling laser.

\section{Theory and Method}
In this section, we introduce the theoretical treatment of micromotion in a linear Paul trap
and the concept of our micromotion minimization technique.

We provide an overview of the trapped single ion motion in a conventional linear Paul trap.
Micromotion in a linear Paul trap arises in a radial plane in an ideal case. 
Therefore, we describe the ion motion only in a radial plane for brevity.
An equation of motion of a single ion that has mass $m$ and charge $Q$ in a linear Paul trap can be described as
\begin{equation}\label{eq_of_motion}
	\ddot{r}_{i} + \left[ a_i + q_i\cos \left( \Omega t \right) \right] \frac{\Omega^2}{4}r_i = \frac{Q E_{{\rm{stray}}, i}}{m}.
\end{equation}
where $t$ is time, $r_i$ is the ion position, and $i = x', y'$ indicates the Cartesian coordinate in the radial plane.
The $a_i$ and $q_i$ are usually called the trapping parameters and
can be determined by the mass of the ion, the ion trap geometric condition, and the applied voltage to the electrodes.
${{\bm E}}_{{\rm{stray}}} = \left( E_{{\rm stray}, x'}, E_{{\rm stray}, y'} \right)$ is the stray electric field and 
is assumed to be a static and uniform electric field in the region of the trapped ion.

The solution of Eq.(\ref{eq_of_motion}), which is called the Mathieu equation, in the case where $|a_i|, q_i^2 \ll 1$ is approximately
\begin{equation}\label{motion}
	r_i(t) \approx \left[ r_{0,i} + r_{1, i}\cos \left( \omega_i t + \phi_i \right) \right] \left[ 1+\frac{q_i}{2}\cos \left(\Omega t\right) \right],
\end{equation}
where 
\begin{equation}\label{r0}
	r_{0, i} \approx \frac{Q E_{{\rm{stray}}, i}}{m {\omega_i}^2}.
\end{equation}
Here, $r_{1, i}$ and $\phi_i$ are determined by the initial condition of the ion.
The secular frequency $\omega_i$ is described as $\omega_i = \frac{\Omega}{2}\sqrt{a_i+{q_i}^2/2}$.

From Eq.(\ref{motion}), the ion motion is characterized by two frequencies:
one is that of the motion in a harmonic potential $\omega_{i}$ commonly called secular motion,
and the other is the micromotion following the RF frequency $\Omega$.

The term $r_{0, i}$ in Eq.(\ref{motion}) that is dependent on the stray electric field shows
the equilibrium position of the ion from the RF node.
In the case of stray electric field $E_{{\rm{stray}}, i} = 0$, a trapped ion is confined at the RF node.
In such a well-compensated condition, the ion's motional energy can be minimized by the laser cooling technique.
A non-negligible stray electric field shifts the ion trapping position given by Eq.(\ref{r0}) from the RF node.
This positional shift arises from a motion driven by the RF electric field, which is called excess micromotion.
This excess micromotion cannot be significantly reduced by laser cooling
because it synchronizes with the RF period, and energy transfer between the ion and RF electric field always exists in a single RF period\cite{doi:10.1063/1.367318}.
Thus, to minimize the ion motional energy in the ion trap,
the external electric field should be applied to reduce the stray electric field and
compensate for the trapping position of the ion in two dimensions.

Excess micromotion can arise not only from a finite stray electric field, such as one caused by a charged insulator, for example,
but also from RF phase mismatching between the electrodes
or geometric distortion of the ion trap.
However, we will not consider these factors here.

Next, we discuss the trapping position shift of the ion from the RF node due to the stray electric field in relation to trap confinement.
From Eq.(\ref{r0}), the ion's trapping position is described by an equilibrium of the spring force resulting from the pseudopotential and the electrostatic force of the stray field.
Therefore, the ion trapping position is close to the RF node with asymptotically increasing RF confinement.
In case of infinite trap confinement $\omega_{i} \rightarrow \infty$,
the trapping position of the ion is coincident with the RF node under the condition of a finite stray electric field $E_{{\rm{stray}}, i} \neq 0$.

Let us assume two different trap confinements $\omega_i$ and $\omega'_i$.
Then the ion displacement between the two different trapping positions $\Delta r_{0, i}$ can be written as
\begin{equation}\label{delta}
	\Delta r_{0, i} = \left| \frac{Q E_{{\rm{stray}, i}}}{m}\left( \frac{1}{\omega_i^2} - \frac{1}{{\omega'_i}^2} \right) \right|.
\end{equation}

This equation shows that the stray electric field is detectable by 
measuring the ion displacement due to the two different trap frequencies.
The ion displacement is proportional to the stray electric field;
thus, the ion displacement becomes zero
when the stray electric field is reduced by applying another external electric field for compensation.

The ion displacement in Eq.(\ref{delta}) is represented in one dimension.
Therefore, $\Delta r_0 = \sqrt{ \sum_i{\Delta r_{0, i}}^2}$ yields the ion displacement in the radial plane.

\begin{figure}[tb]
\includegraphics[clip, width = 8.5cm]{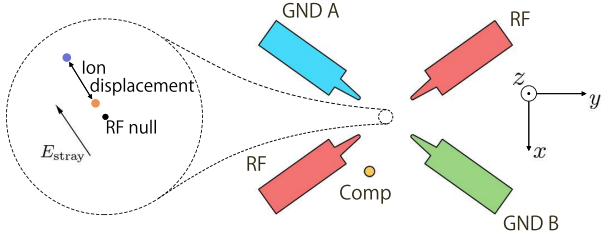}
\caption{\label{fig:fig1}
Schematic diagram of ion displacement in the radial plane (left side)
and the linear Paul trap seen from the axial direction in our system (right side, not scaled).
(left) An ion in a linear Paul trap trapped at the RF node without a stray electric field.
The ion's equilibrium position is shifted proportionally to a finite stray electric field (orange point).
 Furthermore, the shift amount of the ion also depends on its radial confinement.
If the trap frequency is decreased by tuning the RF power, the equilibrium position shifts away from the RF null (purple point).
Excess micromotion can be detected by the ion displacement between the two different confinements that rely on the stray electric field amplitude.
(right) An ion is confined by RF voltage applied to the RF electrodes and RF ground (GND) electrodes, in two dimensions, and
axial confinement is achieved by DC voltage applied to the END electrodes located on the front and back of the GND electrodes.
Minimization of micromotion is carried out by applying DC voltage to the GND A and Comp electrodes to cancel out the stray electric field (GND B electrode is connected to laboratory ground).
}
\end{figure}

Our compensation method concept is based on measuring the ion displacement due to the difference in the ion confinement in the radial plane (see the left side of Fig\ref{fig:fig1}).
Moreover, we determine the trap condition in which the trapping position does not change with RF power variation by applying an external electric field.

An external force can also cause the trapping position to shift
due to the light pressure of the Doppler cooling laser or gravity, for example.
These external forces can cause systematic errors in micromotion compensation. Therefore we have to manage these forces carefully for the precise detection of the RF node.                                                                                                                                                                                                                                                                                                                                                                                                                     
On the other hand, this fact shows that the detection of the trapping position 
can also be used as a small force sensor.

\section{Experimental Setup}
In this study, we used a conventional linear Paul trap
that has nine electrodes (two RF electrodes, two RF ground (GND) electrodes,
four END electrodes, and a Comp electrode for compensation of ion position).
The schematic diagram of our linear ion trap seen from the axial direction is shown in the right side of Fig. \ref{fig:fig1}.
Radial confinement was realized by two RF electrodes and two GND electrodes (shown as RF, GND A, and GND B in Fig.\ref{fig:fig1}).
DC voltage was applied to four END electrodes, which are attached to the front and back of the GND electrodes along the $z$ direction
to achieve axial trap confinement.

Neutral ytterbium vapor from an atomic oven was photoionized by $399~\rm{nm}$ and $369~\rm{nm}$ lasers
and then a single  ion was trapped at the center of the ion trap.
The trapped single $\rm{^{174}Yb^+}$ ion was cooled by Doppler cooling of the $S_{1/2}$--$P_{1/2}$ transition 
using a $369~\rm{nm}$ cooling laser and a $935~\rm{nm}$ repumping laser.
We also irradiated $760~\rm{nm}$ laser to pump out the $\rm{Yb^+}$ ion in $F_{7/2}$ state to the cooling cycle.
These lasers were overlapped and co-propagated through the trap center along the axial direction.
Another set of cooling and repumping lasers was incident through the trap center in the $y$-$z$ plane
to cool the radial ion motion efficiently.

In typical trapping conditions, we drove the RF electrodes using a helical resonator with $570~\rm{V}$ 
at a resonance frequency of $\Omega = 2\pi\times 24.7~\rm{MHz}$.
The radial secular trap frequency was $\omega_r = 2\pi\times (1.90~\rm{MHz}, 1.75\rm{MHz})$.
We also applied $68.5~\rm{V}$ to the END electrodes, and the secular frequency along the $z$ direction was 
$\omega_z = 2\pi\times1.17~\rm{MHz}$.
We deliberately applied modulation voltage superposing to the GND A or END electrode and
excited the secular ion motion to independently measure the trap frequency.
In our proposed micromotion minimization method, we trapped the ion in several radial trap confinement conditions
by changing the RF power applied to the helical resonator.

To compensate the excess micromotion in the radial plane, we applied  additional compensation voltages $V_{\rm{GND}}$ and $V_{\rm{Comp}}$ to the electrodes GND A and Comp (colored blue and yellow in Fig.\ref{fig:fig1}), respectively.
The GND A and Comp electrodes were aligned from the trap center at $144.4^{\circ}$ counterclockwise with respect to $y$ axis and
at $16.3^{\circ}$ clockwise with respect to the $x$ axis, respectively.
Each electrode and the trap center made an angle of $109.4^{\circ}$.
Applying DC voltages to these electrodes created a static and approximately uniform electric field around the trap center in each direction.

The ion fluorescence from $S_{1/2}$--$P_{1/2}$ transition due to the laser cooling cycle was collected by an objective lens installed at the top of the ion trap in Fig.\ref{fig:fig1}.
Because our imaging system captured the fluorescence image in the $y$--$z$ plane,
we estimated the ion position in  direction by imaging the lens position when an ion image was in focus\cite{PhysRevA.92.043421}.
Ion images were acquired around the focus position $\pm 20~{\rm{\mu m}}$ typically
while scanning the imaging lens mounted on the motorized stage.
The objective lens for imaging the ion, which had a numerical aperture of 0.28, was mounted on a piezo motor stage and the focusing lens was positioned along the $x$ direction.
The gathered fluorescence light propagated along the $x$ axis and
reached an electron-multiplying charge-coupled device (EMCCD) camera and a photomultiplier tube (PMT).
The imaging magnification of our imaging system was 27.0, and
the estimated collection efficiency of emitted photons from the ion was $1.9\%$.
Finally, approximately $30\%$ of the photons collected by the lens could reach the EMCCD camera.

\section{Result}
\subsection{Measurement of ion position in radial plane}
\begin{figure}[tb]
\includegraphics[clip, width = 8.5cm]{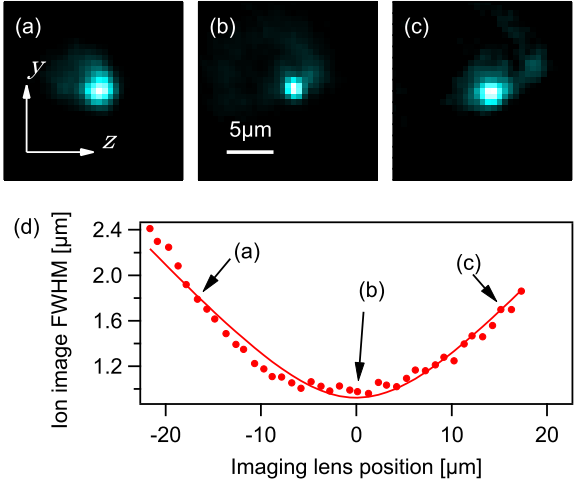}
\caption{\label{fig:fig2}
Ion position measurement in the radial plane by fluorescence image of an ion and its focus position.
(a--c) Ion fluorescence images at different imaging lens positions around the focus position.
We fit this fluorescence image by Gaussian distribution and evaluated the full width at half-maximum (FWHM) in each image.
The coordinates indicated in (a) show the axes that we defined and illustrated in Fig.\ref{fig:fig1}.
The white bar in (b) shows $5~{\rm{\mu m}}$ scale of the images.
(d) Each plot shows the FWHM of the ion image at each objective lens position.
We determined the ion position along the imaging depth (namely $x$) direction by Gaussian beam waist fitting (indicated by the red line).
}
\end{figure}

To compensate for the excess micromotion caused by a stray electric field
based on ion displacement due to trap confinement variation,
we measured the two-dimensional ion position in the radial plane, namely the $x$-$y$ plane in our system.
Figure \ref{fig:fig2} shows a typical result of ion position measurement.
Each fluorescence image in Fig.\ref{fig:fig2}(a--c) was captured during $400~\rm{ms}$ accumulation of emitting photons from an ion;
furthermore, it was fitted by Gaussian to estimate the position in $y$ direction and width of the ion image.
The width of the ion image as a function of imaging lens position is illustrated in Fig.\ref{fig:fig2}(d).
We determined the focus position by fitting these data using an equation of the Gaussian beam waist
and deducing the position in the $x$ direction.
The absolute position of an ion in the radial plane was measured using this procedure.

The absolute position in the $x$ direction had a few micrometer systematic error in each scanning sequence,
caused by lack of piezo motor stage position repeatability.
To avoid this crucial error, we measured the ion position in different radial confinements in a single scan of the stage.

\subsection{Two dimensional ion displacement}
We measured the trapping position of an ion at two different radial trap confinements using the method described in Section II. 
These two equilibrium points yielded the ion displacement at the compensation condition derived from the voltage applied to electrodes GND A and Comp.
To detect and minimize micromotion, we scanned the compensation voltage two-dimensionally
and measured the ion displacement.

\begin{figure}[tb]
\includegraphics[clip, width = 6.5cm]{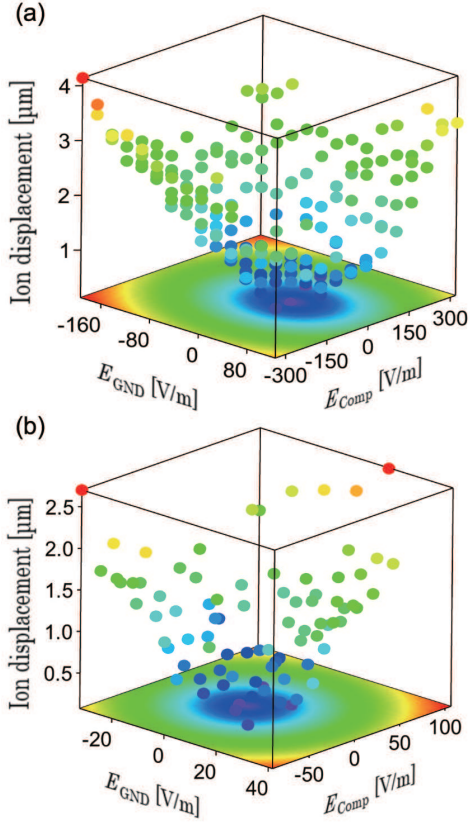}
\caption{\label{fig:fig3}
Measured ion displacement with various combinations of electric fields derived by compensation voltage applied to the GND A and Comp electrodes.
Each point represents the measured data, and the intensity plot on the bottom shows the fitting of these data by the cone function.
Results of tuning of the RF power variation (a) $3.71~{\rm{dBm}}$ and (b) $4.75~{\rm{dBm}}$.
The obtained residual stray electric field uncertainty is $\left( \Delta E_{\rm{GND}}, \Delta E_{\rm{Comp}} \right) =\left( 3.5~\rm{V/m}, 9.4~\rm{V/m} \right)$ from (b).
The sensitivity of this compensation method depends on the magnitude of the RF power variation
because an ion's equilibrium position can easily shift in the case of a small stray electric field with a low trap frequency.
}
\end{figure}

Figure \ref{fig:fig3} shows ion displacement measurement results for determining the optimum trap condition.
$E_{\rm{GND}, }$, $E_{\rm{Comp}}$ axes show the induced electric field around the small region of the trap center
derived by applying a voltage to the GND A and Comp electrode, respectively.
This electric field for compensation was calibrated from the other measurements by detecting the equilibrium position with the varying voltage applied to the GND A or Comp electrode.
Each point represents the observed ion displacement at each compensation electric field combination.
This ion displacement was induced by varying the radial trap frequency with a finite stray electric field condition. Therefore, we realized a change in the trap frequency by tuning the RF power applied to the RF electrodes.

The ion displacement measurement with $3.71~\rm{dBm}$ RF power variation is shown in Fig.\ref{fig:fig3}(a).
This RF power change corresponds to $(1.90~\rm{MHz}, 1.75~\rm{MHz}) \rightarrow (1.21~\rm{MHz}, 1.09~\rm{MHz})$ variation of the radial trap frequencies. 
This result clearly shows the cone shape expected from Eq.\ref{delta} and 
an intensity profile on the bottom of the graph displays a fitted outcome by a cone function.
We estimated the residual stray electric field uncertainty of $\left( \Delta E_{\rm{GND}}, \Delta E_{\rm{Comp}} \right) =\left( 11.4~\rm{V/m}, 16.5~\rm{V/m} \right)$ from this measurement.

To reveal the optimum trap condition more precisely,
we investigated the same measurement with a larger RF power variation of $4.75~\rm{dBm}$ with smaller scanning of the electric field for compensation, 
and the result is displayed in Fig.\ref{fig:fig3}(b).
In this measurement, we changed the radial confinement of $(1.90~\rm{MHz}, 1.75~\rm{MHz}) \rightarrow (1.06~\rm{MHz}, 0.95~\rm{MHz})$ to evaluate the ion displacement.
$\left( \Delta E_{\rm{GND}}, \Delta E_{\rm{Comp}} \right) =\left( 3.5~\rm{V/m}, 9.4~\rm{V/m} \right)$ was obtained from this result.
The residual stray electric field uncertainty improved compared to the result of the RF power variation $3.71~\rm{dBm}$
because the ion displacement $\Delta r_0$ becomes sensitive by tuning the magnitude of the RF power variation.

This tunable sensitivity of micromotion compensation is a notable point of this method
because we can seek a compensated condition step by step to suit the current trap situation.
Therefore, we expect that a better residual electric field can be obtained by decreasing the lower limit of radial confinement.
However, this is restricted by the instability of an ion with less than $1~\rm{MHz}$ of radial trap confinement in our system.
This trap instability could be caused by a radial trap frequency lower than the axial one,
micromotion induced by other factors (e.g., geometrical distortion of an ion trap),
micromotion in the axial direction due to the breakdown of an ion trap translational symmetry, or a combination of these elements.

The optimum condition of the ion trap can be determined instinctively and simply from Fig.\ref{fig:fig3}
because the minimum ion displacement condition is represented in the two-dimensional parameter space
that corresponds to electric fields canceling out the stray electric field in the two-dimensional radial plane.
This easy-to-recognize representation is one of the advantages of this compensation technique compared to the other techniques based on ion position sensing.
Moreover, only an imaging system of ion fluorescence is required to introduce this compensation measurement.
This convenience is another advantage of this technique.
From the point of view of sensitivity, the obtained uncertainty of the residual electric field is not smaller than but comparable to that of other reported techniques\cite{doi:10.1063/1.4930037}.

\begin{figure}[tb]
\includegraphics[clip, width = 6.5cm]{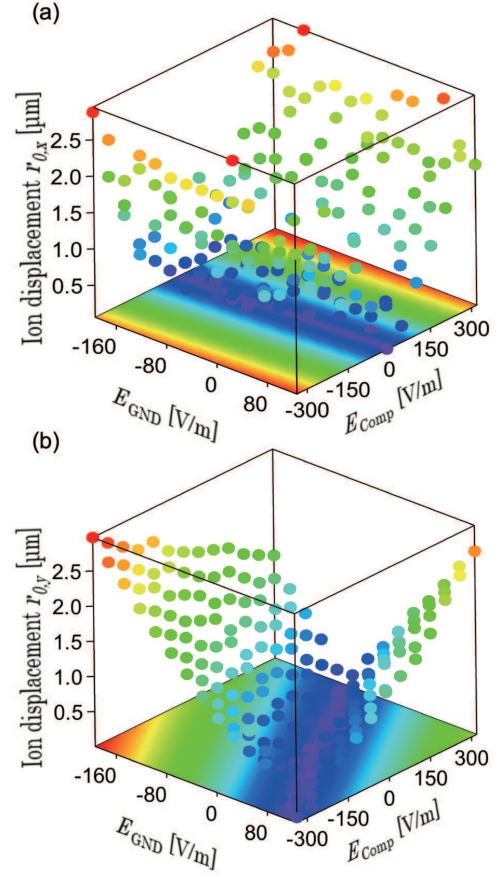}
\caption{\label{fig:fig4}
Measured ion displacement in one dimension with various electric field combinations of $\left( E_{\rm{GND}}, E_{\rm{Comp}} \right)$.
The measured ion displacements induced by an RF power variation of $3.71~{\rm{dBm}}$ is shown as the plotted points and the fitting result of these ion displacement data by a valley shape function is indicated as the intensity plot at the bottom.
We investigated (a) $\Delta r_{0, x}$ and (b) $\Delta r_{0, y}$, namely the ion displacement in $x$ and $y$ directions.
The valley bottoms of (a) and (b) compose lines in $E_{\rm{GND}}$-$E_{\rm{Comp}}$ plane
and the condition of two-dimensional micromotion compensation in Fig.\ref{fig:fig3} is achieved by the intersection of these lines.
}
\end{figure}

Because the ion displacements were evaluated by two absolute equilibrium points of an ion in the radial plane,
we can analyze the ion displacements in the respective axes of $\Delta r_{0, x}$ and $\Delta r_{0, y}$, which are shown in Fig.\ref{fig:fig4}(a) and (b), respectively.
The measured data are indicated as the plots, and the intensity plot at the bottom represents fitting by a valley shape function.
The valley bottoms of the results in Fig.\ref{fig:fig4}(a) and (b) show the micromotion-minimized position in the $x$ and $y$ directions, respectively.
This line represents the nodal line of the RF electric field near the trap center along the $x$ or $y$ direction, and the geometry of the ion trap determines the slope of the line.
Because the valley bottom of each direction only has a sensitivity for the respective axis,
the intersection of these two lines coincides with the two-dimensional micromotion-minimized condition of Fig.\ref{fig:fig3}.
A similar situation occurs in the RF-photon correlation method, which is only sensitive to micromotion in a projection direction to the radial plane of the incident cooling laser.

The data of $\Delta r_{0, x}$ are noisier than the data of $\Delta r_{0, y}$
by comparison between Fig.\ref{fig:fig4}(a) and (b).
This is caused by the low accuracy of ion position determination in the imaging depth ($x$) direction compared to that in the imaging plane ($y$ and $z$ directions).
Because an angle between ${{\bm E}}_{\rm{Comp}}$ near the trap center and the $x$ axis is supposed to be small from the trap geometry,
the shifts of the equilibrium positions induced by $E_{\rm{Comp}}$ are nearly along $x$ direction.
It is reasonable that the value of the residual stray electric field uncertainty in $E_{\rm{Comp}}$ direction
is larger than that in the $E_{\rm{GND}}$ direction. 

\subsection{Limitation of detecting residual micromotion}

Next, to reveal the performance of this measurement method,
the sensitivity and the limitation of detecting residual micromotion are discussed.
We also mention the equilibrium position shift due to forces other than the electric field.
These unexpected shifts may induce systematic error for micromotion minimization
in the case of the compensation method based on ion position detection.

\begin{figure}[tb]
\includegraphics[clip, width = 8.5cm]{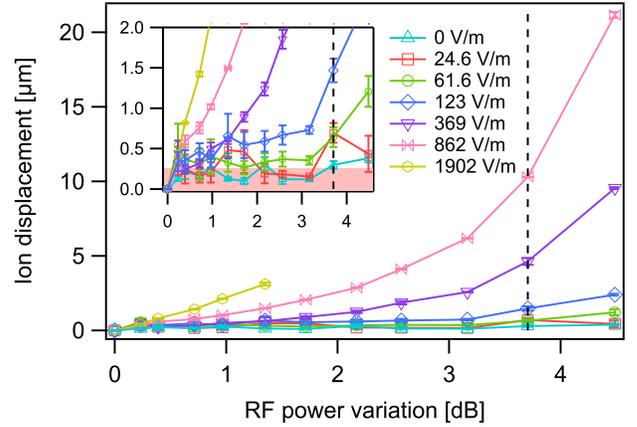}
\caption{\label{fig:fig5}
RF power variation dependence of the ion displacements.
We measured the ion displacements with decreasing trap frequency of the lower side in various static electric fields by tuning ${{\bm E}_{\rm GND}}$.
Each point represents the average of three independent measurements, and the error bar shows the standard error.
The black dashed line indicates the same measurement conditions as Fig.\ref{fig:fig3}(a).
The inset displays an enlarged vertical axis, and
the red shaded area represents the typical uncertainty of an ion displacement measurement.
}
\end{figure}

We evaluated the magnitude of the RF power variation dependence of the two-dimensional ion displacement of $\Delta r_0$ with the various static electric fields in Fig.\ref{fig:fig5}.
In this measurement, to change the magnitude of the RF power variation, the higher trap frequencies were fixed, and only the lower frequencies were tuned.
We started measuring the ion displacements at RF null in the radial plane determined by the procedure discussed above (data set of $0~{\rm V/m}$)
and applied the electric field by tuning ${{\bm E}_{\rm GND}}$.
The RF power variation corresponding to the condition in Fig.\ref{fig:fig3}(a) is shown as the black dashed line
and the inset displays a small region of the vertical axis.

The ion displacements increased along the horizontal axis
because the large equilibrium position shift was derived by low trap confinement.
This result shows the sensitivity of this method in terms of distinguishable residual micromotion at each RF power variation in our system.
It is evident that a better residual stray electric field uncertainty can be achieved by lowering the trap confinement for measuring the ion displacement.
The red shaded area in the inset indicates the uncertainty of the ion displacement determination measurement.
Therefore, the plots in the shaded area cannot be distinguished from one another.
The accuracy of the ion position measurement is mainly determined by a focal point measurement (namely, a position measurement in the $x$ direction) in the current situation
because the typical uncertainty of a focal point is more than one order of magnitude larger than that in the imaging plane. The uncertainty of a focal point depends on the depth of the field.
Thus, we can improve the accuracy of a focal point measurement by replacing the imaging lens with one having a smaller focal length and larger numerical aperture. In principle, this is restricted by the Rayleigh length.
On the other hand, the limitation of ion position determination in the imaging plane (namely, the $y$ and $z$ directions) depends on the imaging resolution.
The calculated localized area of an ion in the ion trap is much smaller than the ion image width at the focal point (typical value is about $1~{\rm{\mu} m}$; refer to Fig.\ref{fig:fig2}(d)).
Therefore, the current imaging resolution is approximately $1~{\rm{\mu} m}$, and it can be reduced approximately by a factor of $0.37$ in terms of the diffraction limit.
To sum up, measuring ion displacement with a considerable RF power variation and accuracy of the ion position determination
are crucial for this micromotion minimization technique.

Finally, we mention the equilibrium position shift of an ion due to forces other than the electric field.
The detectable minimum ion displacements of $\Delta r_{0}$, $\Delta r_{0, x}$, and $\Delta r_{0, y}$ by our imaging system are approximately $0.25~{\rm \mu m}$, $0.25~{\rm \mu m}$, and $0.011~{\rm \mu m}$, respectively (the minimum detectable $\Delta r_{0}$ corresponds to the red shaded area in the inset of Fig.\ref{fig:fig5}).
This detectable minimum ion displacement is equivalent to an ion displacement due to a force of $2.9~{\rm aN}$, $2.9~{\rm aN}$, $125~{\rm zN}$ at $1~{\rm MHz}$ trap frequency.
This detection limit is one order of magnitude larger than that reported in the previous work\cite{blums2018single}.
We do not have to consider the shift in ion equilibrium position due to gravity
because the gravitational force for a single ytterbium ion is $2.8~{\rm yN}$.
On the other hand, we must consider the light pressure due to the cooling laser.
The calculated light pressure is about $100~{\rm zN}$ 
assuming the $S$--$P$ transition of $\rm Yb^{+}$ ion, in which the cooling laser is half red detuned of the line width from the resonance and the saturation parameter is equal to ten.
Therefore, we can ignore the ion position shift due to the cooling laser along the $x$ direction. In contrast, the shift in the $y$ direction is comparable to the accuracy of the ion position determination in the imaging plane.
This can cause a systematic error in the determination of ion position if the cooling laser intensity or detuning fluctuates.

\section{Conclusion}
In conclusion, we have developed and demonstrated a simple method for minimizing the excess micromotion in a linear Paul trap
based on measuring the ion displacement due to a stray electric field.
The magnitude of the stray electric field was evaluated by its ion displacement, which was measured by determining the equilibrium position with two different radial trap frequencies detected by fluorescence images of an ion.
We found the micromotion-compensated condition by searching for an ion displacement of zero, and
the uncertainty of the residual electric field reached a few volts per meter.
The sensitivity achieved is not superior but comparable to that of existing compensation techniques\cite{doi:10.1063/1.4930037}.
Because the equilibrium position shift increases with decreasing trap frequency,
the sensitivity of micromotion detection can easily be tuned using this method.
Furthermore, this compensation technique does not depend on the orientation and detunig of the incident cooling laser or imaging direction.
Therefore, this method is applicable to planar ion traps, stylus ion traps, which limit the propagation angle of lasers, or miniaturized ion traps and systems.
Moreover, complex laser systems or instruments are not required in this measurement
because this method is based on the detection of ion positions with trap potential modulation.
Thus, this technique can simplify the ion trap system and its experiments. 
This can lead to the development of quantum information and sensing techniques using ion traps.

\section{Acknowledgement}
This work was supported by JST-Mirai Program Grant Number JPMJMI17A3, Japan.

\section{Data Availability}
The data that support the findings of this study are available from the corresponding author upon reasonable request.

\bibliography{bib}

\end{document}